# MedSTS: A Resource for Clinical Semantic Textual Similarity

Yanshan Wang[*], Naveed Afzal[*], Sunyang Fu, Liwei Wang, Feichen Shen, Majid Rastegar-Mojarad, Hongfang Liu[†]

Department of Health Sciences Research

Mayo Clinic, Rochester, MN

[*]Co-first authors
[†]Corresponding author:
Email: Liu.hongfang@mayo.edu

# Abstract

The wide adoption of electronic health records (EHRs) has enabled a wide range of applications leveraging EHR data. However, the meaningful use of EHR data largely depends on our ability to efficiently extract and consolidate information embedded in clinical text where natural language processing (NLP) techniques are essential. Semantic textual similarity (STS) that measures the semantic similarity between text snippets plays a significant role in many NLP applications. In the general NLP domain, STS shared tasks have made available a huge collection of text snippet pairs with manual annotations in various domains. In the clinical domain, STS can enable us to detect and eliminate redundant information that may lead to a reduction in cognitive burden and an improvement in the clinical decision-making process. This paper elaborates our efforts to assemble a resource for STS in the medical domain, MedSTS. It consists of a total of 174,629 sentence pairs gathered from a clinical corpus at Mayo Clinic. A subset of MedSTS (MedSTS_ann) containing 1,068 sentence pairs was annotated by two medical experts with semantic similarity scores of 0-5 (low to high similarity). We further analyzed the medical concepts in the MedSTS corpus, and tested four STS systems on the MedSTS_ann corpus. In the future, we will organize a shared task by releasing the MedSTS_ann corpus to motivate the community to tackle the real world clinical problems.

# 1. Introduction

The wide adoption of electronic health records (EHRs) has provided a way to electronically document a patient's medical conditions, thoughts, and actions among the care team (Blumenthal 2011, Williams, Mostashari et al. 2012). While the use of EHRs has led to an improvement in quality of healthcare, it has introduced new challenges (Kuhn, Basch et al. 2015). One such challenge, ironically, stems from the ease of use of EHRs; the growing use of copy-and-paste, templates, and smart phrases causes clinical notes to bloat in size with poorly organized or erroneous documentation (Embi, Weir et al. 2013, Zhang, Pakhomov et al. 2014). EHRs are effectively optimized to store massive amounts of information at the cost of adding to the cognitive burden of tracking multiple complex medical problems or maintaining continuity and quality of the clinical decision-making process.

As such, there is a growing need for automated methods to better synthesize patient data from EHRs and reduce the cognitive burden in clinical decision-making process for providers. Patient data can be scattered in several heterogeneous sources. Tools are desired that can aggregate data from diverse sources, minimize data redundancy, and organize and present the data in a user-friendly way to reduce the cognitive burden (Schiff and Bates 2010). Previous studies have used different automated methods for identification of redundant/new relevant information from both inpatient and outpatient notes (Wrenn, Stein et al. 2010, Zhang, Pakhomov et al. 2011, Zhang,

Pakhomov et al. 2014). For example, Zhang et al. (Zhang, Pakhomov et al. 2014) used statistical language models to identify relevant new information from patient's progress notes. Evaluation of their methods against expert-derived gold standards found that clinical notes contained 76% redundant information. The best method was able to attain a precision of 0.74, recall 0.83 and F-score of 0.78 in identifying new information in inpatient notes. Clinical text summarization focuses on collecting and synthesizing important patient information for the purpose of facilitating healthcare professionals to perform a wide range of clinical tasks efficiently (Friedman and Elhadad 2014, Hirsch, Tanenbaum et al. 2015). It presents a different set of challenges from general text summarization, such as information redundancy, temporality, complexity of medical terminologies and missing data (Pivovarov and Elhadad 2015), Automatic clinical text summarization becomes more necessary for transferred patients since they usually bring a overwhelmingly large number of digitally-faxed scanned or hand-carried outside materials that it would be impossible for a practitioner to read during a regular medical visit (Moon, Liu et al. 2017, Pivovarov and Elhadad 2015).

One enabling technique for automatically summarizing information is to compute semantic similarity between text snippets and remove highly similar text snippets. In the general English domain, the SemEval Semantic Textual Similarity (STS) shared tasks (Agirre, Diab et al. 2012, Agirre, Cer et al. 2013, Agirre, Banea et al. 2014, Agirrea, Baneab et al. 2015, Agirrea, Baneab et al. 2016) have been organized since 2012 to motivate the natural language processing (NLP) community to develop automated methods for this requirement. In the medical domain, however, there are few STS systems developed for computing clinical text similarity. The main reason is the lack of clinical STS resources for NLP researchers. To bridge the gap, we describe our effort in creating an STS resource, called MedSTS dataset, consisting of sentence pairs extracted from our clinical corpus at Mayo Clinic. We selected unique sentences and made sentence pairs using various surface similarity measures. After generating sentence pairs, two medical experts with clinical background were asked to annotate a subset of MedSTS (MedSTS_ann) with semantic similarity scores of 0-5 (low to high similarity), which could later be used as the gold standard. Based on the MedSTS_ann, we plan to organize a shared medical STS task akin to SemEval STS shared task that motivates the community to tackle the real clinical practical problem. Since clinical text contains highly domain-specific terminologies (Meystre, Savova et al. 2008, Pradhan, Elhadad et al. 2014)., participant STS systems will also be tailored and designed differently from those in general domain, which will be our main contribution to both NLP and clinical community.

This paper is structured as follows. We first provide background information regarding STS and its use in various NLP applications. The methods adopted for generating the STS resource are presented in Section 3. We then present an overview of the STS resource in Section 4 and discuss potential clinical NLP applications in Section 5.

## 2. Background

Semantics is a study of the meaning of natural language expressions and the relationships between them. In computational semantics, we focus on automatically constructing and reasoning with the meaning of natural language expressions (Mitkov 2005). Semantic textual similarity (STS) assessment is a common task in computational semantics aiming to calculate the similarity between natural language expressions, e.g., sentences or text snippets, on the basis of their semantic meaning or content. STS is closely related to paraphrase detection and textual entailment tasks (Majumder, Pakray et al. 2016). STS produces a scaled output to show how similar two text snippets are. STS is a challenging task as the same idea (semantic meanings) can easily be articulated in numerous different ways and the same set of words can be combined into different sentences with completely different semantic interpretations.

STS is an integral part of many NLP applications such as information retrieval (Rada, Mili et al. 1989, Srihari, Zhang et al. 2000), word sense disambiguation (Patwardhan, Banerjee et al. 2003), question answering (Tapeh and Rahgozar 2008), automatic machine translation evaluation (Kauchak and Barzilay 2006), recommender system (Blanco-Fernández, Pazos-Arias et al. 2008), information extraction (Atkinson, Ferreira et al. 2009) and textual summarization (Aliguliyev 2009). Automated extraction from narrative clinical notes has played an important role in meaningful use of EHRs for clinical and translational research (Wang, Wang et al. 2018). The earliest methods to compute the similarity between two sentences used word-to-word similarity methods (Corley and Mihalcea 2005) computed using measures from the WordNet similarity package (Pedersen, Patwardhan et al. 2004) as well as simple vector space models (Salton, Wong et al. 1975). There are two main resources leveraged for measurement of semantic similarity: massive corpora of text documents (Barzilay and McKeown 2005, Islam and Inkpen 2008) and semantic resources and knowledge bases (Li, McLean et al. 2006, Corley 2007) such as WordNet (Miller 1995) and Wikipedia. Many researchers have used supervised machine learning approaches where multiple similarity measures and features are combined to compute semantic similarity (Bär, Biemann et al. 2012, Šarić, Glavaš et al. 2012).

The SemEval STS shared tasks (Agirre, Diab et al. 2012, Agirre, Cer et al. 2013, Agirre, Banea et al. 2014, Agirrea, Baneab et al. 2015, Agirrea, Baneab et al. 2016) have played a pivotal role in attracting an increasing amount of interest in the NLP community to the question of textual similarity. These STS tasks examined semantic similarity between two sentences using datasets from various domains by assigning a similarity score of 0-5 to each sentence pair on the basis of their semantic equivalence. For shared tasks (Agirre, Diab et al. 2012, Agirre, Cer et al. 2013, Agirre, Banea et al. 2014, Agirrea, Baneab et al. 2015, Agirrea, Baneab et al. 2016), STS sentence pairs were built using various publically available datasets such as the Microsoft Research Paraphrase Corpus (MSR-Paraphrase)[1], the Microsoft Research Video Research Corpus (MSR-Video)[2], machine translation evaluation sentences (SMTeuroparl)[3], sense definition pairs of OntoNotes (Hovy, Marcus et al. 2006), news headlines (Best, van der Goot et al. 2005), image description (Rashtchian, Young et al. 2010), tweet-news pairs (Guo, Li et al. 2013), answers-student pairs (Dzikovska, Moore et al. 2010), answers-forums pairs from the Stack Exchange answers websites[4], and plagiarism corpus (Clough and Stevenson 2011). The performance of participating systems was evaluated using the Pearson correlation coefficient (Pearson 1895) between the system scores and the human scores. The STS shared tasks datasets have been used for various NLP tasks by the research community e.g. to predict alignments and constituents similarities (Li and Srikumar 2016), semantic indexing of multilingual corpora (Raganato, Camacho-Collados et al. 2016), paraphrastic sentence embeddings (Wieting and Gimpel 2017), and automatic evaluation of machine translation metrics (Magnolini, Vo et al. 2016).

---

[1] http://research.microsoft.com/en-us/downloads/607d14d9-20cd-47e3-85bc-a2f65cd28042/
[2] http://research.microsoft.com/en-us/downloads/38cf15fd-b8df-477e-a4e4-a4680caa75af/
[3] http://www.statmt.org/wmt08/shared-evaluation-task.html
[4] http://stackexchange.com/

# 3. Methods

## 3.1 Data Collection

The construction of a dataset by gathering naturally occurring pairs of sentences with different degree of semantic equivalence is a very challenging task in itself. We extracted EHRs data from Mayo Clinic's clinical data warehouse (Wu, Liu et al. 2012). From the data warehouse, we selected unique sentences from 3 million de-identified clinical notes of patients receiving their primary care at Mayo Clinic. In order to obtain the de-identified sentences, we removed protected health information (PHI) by employing a frequency filtering approach (Li, Rastegar-Mojarad et al. 2015) based on the assumption that sentences appearing in multiple patients' records tend to contain no PHI information. This process resulted in 14.9 million unique sentences with 361.9 million tokens. This study has been approved by the institutional review board (IRB).

## 3.2 Sentence Pairs Selection

Following the lead of the SemEval shared tasks; we used the averaged value of three surface lexical similarities as the measurement to find candidate sentence pairs with some level of *prima facie* similarity. First, a sequence-matching algorithm compares the character sequence in one text snippet with that in the other text snippet based on Ratcliff/Obershelp pattern matching algorithm (Black 2004). Specifically, suppose that $|S_1|$ and $|S_2|$ are lengths of strings $S_1$ and $S_2$ respectively and that $K_m$ is the number of matching characters, the similarity between strings $S_1$ and $S_2$ is defined by

$$Sim_{RO} = \frac{2 * K_m}{|S_1| + |S_2|}.$$

Since $K_m \leq |S_1|$ and $K_m \leq |S_2|$ always hold, this algorithm returns a similarity score between 0 and 1, which shows the surface similarity between the two snippets. Second, we computed the cosine similarity between two text snippets. This is a commonly used measurement where text snippets are transformed into a vector space in order to determine similarity between word vectors using Euclidean cosine rule. Suppose that V is a set of unique words occurred in strings $S_1$ and $S_2$. $S_1$ and $S_2$ can be represented in the same vector space as $\mathbf{s}_1$ and $\mathbf{s}_2$ respectively where each component corresponds to the word in V and the value is the word frequency. The cosine similarity between strings $S_1$ and $S_2$ is defined by

$$Sim_{cos} = \frac{\mathbf{s}_1 \cdot \mathbf{s}_2}{\|\mathbf{s}_1\|\|\mathbf{s}_2\|}$$

Third, we used Levenshtein distance, defined as the minimum number of edits required transforming one text snippet into the other. These edit operations are insertion, deletion and substitution of a single character. We divided the Levenshtein distance by the number of characters in the longer string to normalize the result to [0,1], which is denoted as $Sim_{lev}$.

All methods assign a scalar score between a maximum of 1 if two text snippets are identical, and a minimum of 0 for complete difference. We average these three scores to get a final surface similarity score for a given pair of sentence. We did a pairwise comparison of every sentence in the corpus and experimented with different score ranges and empirically selected all sentence pairs where the average score was greater or equal to 0.45. STS shared task (Agirrea, Baneab et al. 2015) has also sampled sentence pairs using different string similarity values based on the nature of the text. This resulted in 174,629 total sentence pairs, which constructs the clinical semantic textual similarity dataset, MedSTS. Fig. 1 shows the distribution of sentence pairs.

In order to build sentence pairs dataset that would reflect a uniform distribution of similarity ranges, we sampled the dataset at certain range (between 0 and 1) of string surface similarity. We randomly selected equal number of sentence pairs from five scales of surface similarity range [0.45 – 0.95] from the dataset resulting in 1,250 sentence pairs overall. This dataset of 1,250 sentences is a subset of MedSTS (denoted as MedSTS_ann) that will be distributed to participants in our future MedSTS shared task.

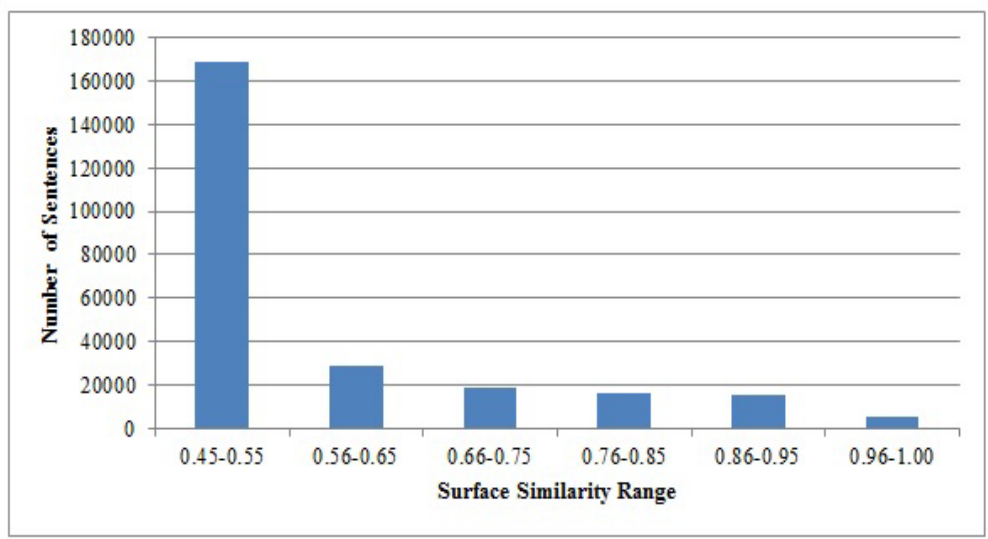

**Fig 1. Sentence pairs distribution on the basis of surface similarity measures**

## 3.3 Annotation

After the sentence pair selection phase, two clinical experts were asked to annotate each sentence pair in the MedSTS_ann on the basis of their semantic equivalence. Both annotators were vastly experienced with many years of experience of clinical domain. Table 1 demonstrates a 6-point ordinal similarity scale along with definitions and examples where a similarity score of 0 denotes complete dissimilarity between two sentences. A similarity score of 1 shows that two sentences are not equivalent but are topically related to each other while similarity score of 2 indicates that two sentences agree on some details mentioned in them. The similarity score of 3 implies that there are some differences in important details described in two sentences while a score of 4 represents that the differing details are not important. The score of 5 represents that two sentences are completely similar.

| Score | Examples |
|---|---|
| 5 | *The two sentences are completely equivalent, as they mean the same thing.*<br><br>S1 → Albuterol [PROVENTIL/VENTOLIN] 90 mcg/Act HFA Aerosol 2 puffs by inhalation every 4 hours as needed.<br><br>S2 → Albuterol [PROVENTIL/VENTOLIN] 90 mcg/Act HFA Aerosol 1-2 puffs by inhalation every 4 hours as needed #1 each. |
| 4 | *The two sentences are mostly equivalent, but some unimportant details differ.*<br><br>S1 → Discussed goals, risks, alternatives, advanced directives, and the necessity of other members of the surgical team participating in the procedure with the patient.<br><br>S2 → Discussed risks, goals, alternatives, advance directives, and the necessity of other members of the healthcare team participating in the procedure with the patient and his mother. |
| 3 | *The two sentences are roughly equivalent, but some important information differs/missing.*<br><br>S1 → Cardiovascular assessment findings include heart rate normal, Heart rhythm, atrial fibrillation with controlled ventricular response.<br><br>S2 → Cardiovascular assessment findings include heart rate, bradycardic, Heart rhythm, first degree AV Block. |
| 2 | *The two sentences are not equivalent, but share some details.*<br><br>S1 → Discussed risks, goals, alternatives, advance directives, and the necessity of other members of the healthcare team participating in the procedure with (patient) (legal representative and others present during the discussion).<br><br>S2 → We discussed the low likelihood that a blood transfusion would be required during the postoperative period and the necessity of other members of the surgical team participating in the procedure. |
| 1 | *The two sentences are not equivalent, but are on the same topic.*<br><br>S1 → No: typical 'cold' symptoms; fever present (greater than or equal to 100.4 F or 38 C) or suspected fever; rash; white patches on lips, tongue or mouth (other than throat); blisters in the mouth; swollen or 'bull' neck; hoarseness or lost voice or ear pain.<br><br>S2 → New wheezing or chest tightness, runny or blocked nose, or discharge down the back of the throat, hoarseness or lost voice. |
| 0 | *The two sentences are completely dissimilar.*<br><br>S1 → The risks and benefits of the procedure were discussed, and the patient consented to this procedure.<br><br>S2 → The content of this note has been reproduced, signed by an authorized |

| | physician in the space above, and mailed to the patient's parents, the patient's home care company. |
|---|---|

Table 1: Similarity scores with explanations and examples

The two annotators made their scoring assessment independently. Finally, similar to the annotation in the SemEval STS shared tasks, we utilized the average of their scores as the gold standard for evaluating STS systems.

# 4. Results

## 4.1 Corpus Analysis

First, we would like to demonstrate the medical concepts covered in the MedSTS dataset. We processed all the sentences in MedSTS using cTAKES (Savova, Masanz et al. 2010) to find information related to the following four main categories of unified medical language system (UMLS)[5] semantic types: sign and symptom, disorder, procedure and medication. Since the UMLS semantic types provide a high-level structure for organizing concepts in the biomedical domain, illustrating the semantic types in the corpus reveals the medical conceptual coverage of the proposed resource. Fig. 2 shows the logarithm of frequencies of each semantic type for both MedSTS and the MedSTS_ann. We found that sign and symptoms (5,299) and disorders (1,222) are mentioned more frequently compared to procedures (634) and medications (41) in MedSTS. Similarly, we found that the MedSTS_ann contains more unique sign and symptoms (334) compare to unique disorders (164), procedures (124) and medications (20). The most frequent categories in each semantic type are consistent for MedSTS and MedSTS_ann. For example, illness, diagnosis, pain and follow-up are the most frequent sign and symptoms while the most frequent disorders include rash, injury, rectal bleeding and side effects. The most frequent procedures include surgical, therapy, respiratory assessment and immunization whereas the most frequent medications include flovent hfa, novolog and epipen. The results in Fig. 2 show that our STS resource provides a wide coverage of the selected UMLS semantic types. Since our previous study validated that the medical concept distributions between the sentences extracted by the frequency-filtering strategy and the entire EHR corpus are similar (Li, Rastegar-Mojarad et al. 2015), the MedSTS dataset could be a representative subset for the EHR corpus.

---

[5] https://www.nlm.nih.gov/research/umls/

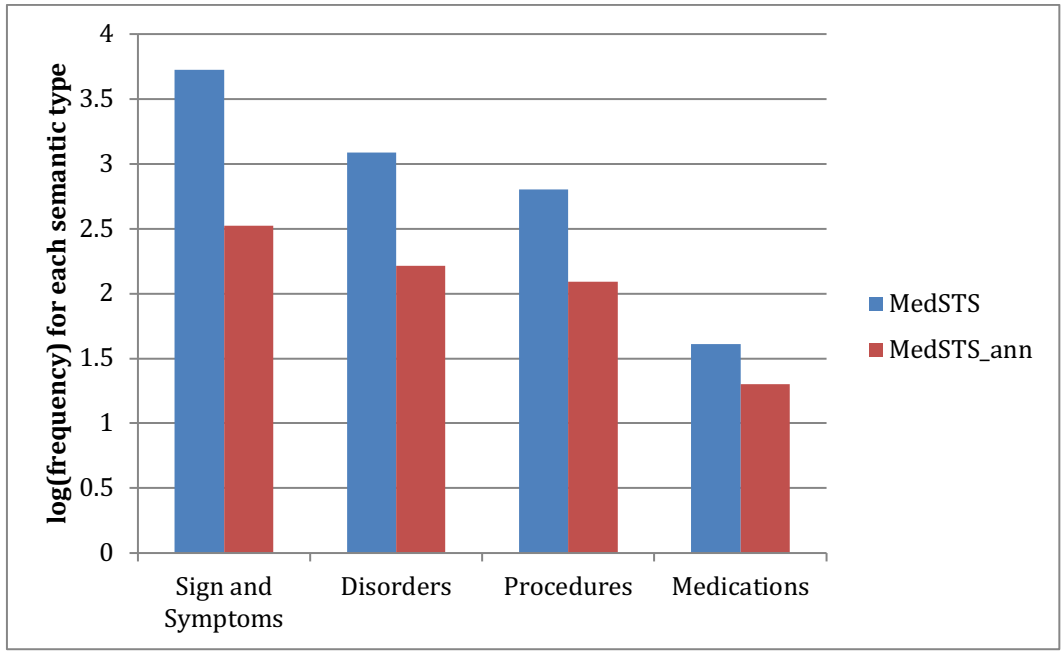

**Fig 2.** Frequencies (log) of frequent UMLS semantic types for each semantic type in the MedSTS and MedSTS_ann datasets.

### 4.2 Annotation Results

The expert annotated clinical STS dataset contained 54,161 word tokens and the average sentence length was 51 words. Fig. 3 shows the distribution of similarity scores assigned to sentence pairs by each annotator. The agreement between the two annotators was high, with a weighted Cohen's Kappa of 0.67.

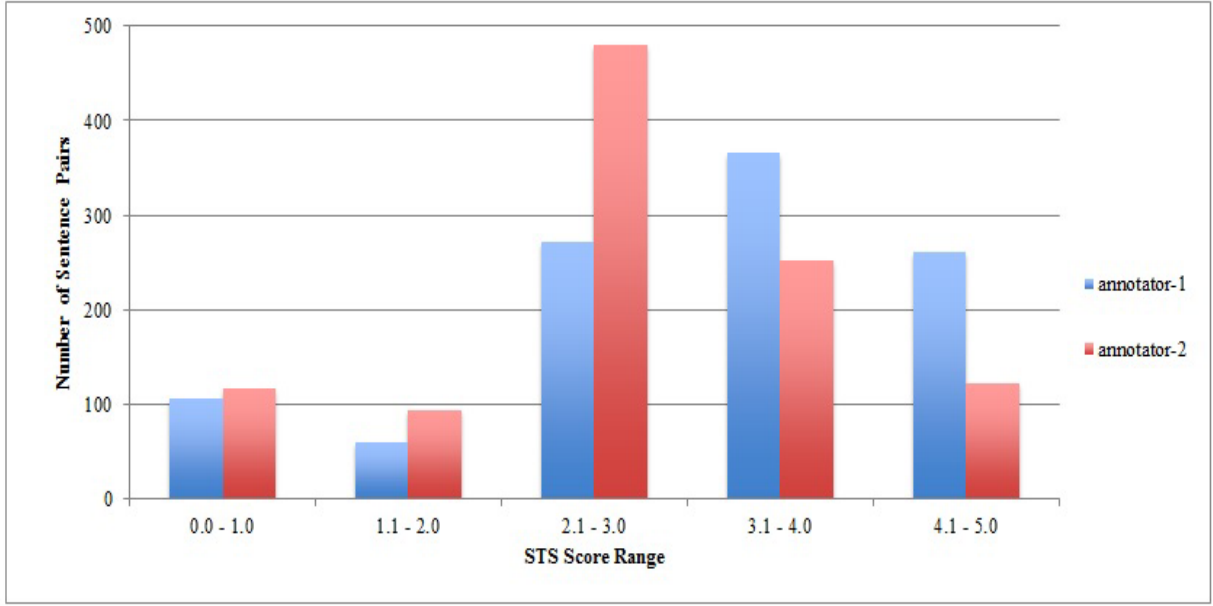

**Fig 3. Annotators' STS score distribution**

### 4.3 Baseline system results

We utilized the aforementioned three surface similarity methods (i.e., Ratcliff/Obershelp's method, cosine similarity, and Levenshtein distance) as well as an ensemble of these methods (the mean of their similarity scores, i.e., $\frac{1}{3}(Sim_{RO} + Sim_{cos} + Sim_{lev})$) as baseline systems. In addition to the MedSTS_ann, four datasets of SemEval-2016 STS task, namely Answers, Headlines, Plagiarism, and Postediting, were utilized to compare the performance of baseline systems on datasets in the general domain with that in the medical domain. The Question dataset from SemEval-2016 was not used since clinical notes in our dataset did not contain question sentences. The system performance is evaluated using the Pearson correlation coefficient between the system scores and the gold standard. Table 2 lists the results of baseline methods on the SemEval-2016 datasets and the MedSTS_ann. We can observe that the performance on MedSTS_ann is inferior to that on the most STS datasets in the general domain for all the baseline systems. This result shows that the clinical MedSTS dataset is more complex than the general domain STS datasets.

| Baseline Method | STS16-Answers | STS16-Headilnes | STS16-Plagiarism | STS16-Postediting | MedSTS_ann |
|---|---|---|---|---|---|
| Ratcliff/Obershelp | 0.5662 | 0.6785 | 0.7084 | 0.8382 | 0.5473 |
| Cosine Similarity | 0.4788 | 0.7436 | 0.7825 | 0.8510 | 0.6109 |
| Levenshtein Distance | 0.5665 | 0.6621 | 0.6901 | 0.8399 | 0.6801 |
| Ensemble | 0.5372 | 0.6947 | 0.7270 | 0.8430 | 0.6128 |

**Table 2**: Pearson correlation coefficient of baseline methods

# 5. Discussion

Redundancy in free text EHRs has become a big challenge for the secondary use of EHRs. According to clinicians (Kuhn, Basch et al. 2015), there is a growing need to improve the clinical documentation process. Copying or importing text from one note to another substantially increases the probability of redundant and erroneous information that can ultimately lead to a clinical error (Singh, Giardina et al. 2013). In a recent study (Wang, Khanna et al. 2017) conducted at the University of California San Francisco Medical Center, over 23,000 progress notes were reviewed over an eight-month period. In this study, they found that 46% of text in each progress note was copied. The application of NLP methods to address this challenge has not been fully explored, mainly due to the limited access of data caused by patient privacy and data confidentiality constraints. In this paper, we aim to bridge the gap by creating an STS resource consisting of sentence pairs extracted from our clinical corpus at Mayo Clinic. Our proposed resource will motivate researchers to develop NLP systems to reduce EHR redundancy and potentially increase usability, portability, and generalizability of the NLP systems.

The sentences in the proposed MedSTS dataset were extracted from actual clinical notes at Mayo Clinic. We asked two clinical experts to annotate the similarity between the sentence pairs in the MedSTS_ann. The annotated similarity scores could be utilized as the gold standard for evaluating STS systems. The distribution of scores (Fig. 2) can be seen to be approximately normal, which is consistent with the feedback from the annotators that the similarities for most pairs were intermediate. The annotators struggled to make STS decision consistently due to the scoring range [0-5] and there is a need for more definitions and examples added to the annotation guidelines from clinical perspective. SemEval STS shared tasks (Agirre, Diab et al. 2012, Agirre, Cer et al. 2013, Agirre, Banea et al. 2014, Agirrea, Baneab et al. 2015, Agirrea, Baneab et al. 2016) have used multiple annotators and assessed the quality of annotation by measuring the correlation of each annotator with the average of the rest of annotators, and then averaging the results. The other challenge is related to the structure of clinical notes in the Mayo corpus. Our STS corpus was developed with sentences from clinical notes without considering different note types and note section. The UMLS semantic type distribution of the STS resource shows that there are more unique sign and symptoms than unique disorders, procedures and medications. A refinement of the STS resource could extract sentences from specific note types or sections. For example, extracting more sentences about procedures from surgical/therapy notes, and medications from medication section in clinical note. By doing so, the resource will have a balanced quantity of each semantic type and facilitate training process in machine learning techniques.

The experimental comparison of baseline systems on datasets from MedSTS and general domain shows that the clinical STS dataset is more complex than the general domain STS datasets. The reason is that the MedSTS dataset contains many medical terminologies. Determining the similarity between medical terminologies is challenging, particularly in the medical domain, due to the complexity of synonymous medical terms and the hierarchy of medical concepts (Peterson, Pakhomov et al. 2007). Therefore, the STS system for the MedSTS dataset should consider using medical domain-specific thesauri in addition to advanced similarity techniques as in the STS system for general domain dataset.

The ability to organize concepts on the basis of their similarity or relatedness to each other is an essential step in the human mind and in many applications of NLP. STS on a sentence level is a vital feature of automatic text summarization (Ferreira, Lins et al. 2016). STS has been a popular research topic in general domain due to STS shared tasks (Agirre, Diab et al. 2012, Agirre, Cer et al. 2013, Agirre, Banea et al. 2014, Agirrea, Baneab et al. 2015, Agirrea, Baneab et al. 2016) but there is not much work done in clinical domain. There has been comparatively little work on STS between concepts in clinical text and the exploration of such information for the purpose of automated clinical summarization (Pivovarov and Elhadad 2015). In clinical domain, STS can be used in patient cohort identification where a user's query could be mapped to multiple semantically similar equivalent formulations. Moreover, the use of STS can significantly reduce excessive redundant information that results in information overload, cognitive burden and difficulties in effective decision-making process at the point-of-care and there is a growing need for computational methods that can decrease the cognitive load of a clinician and increase healthcare efficacy.

Our work has three limitations. First, the size of the clinical STS resource is relatively small. It is developed using only clinical notes from a single institute. The second limitation is that our annotation schema utilizes the conventional STS annotation guidelines with limited consideration of clinical properties. The third limitation is that only two clinical experts manually annotate the dataset. Annotation of SemEval STS shared tasks was performed using crowdsourcing on Amazon Mechanical Turk, which is not applicable for our dataset due to the sensitive patient data.

In the future, in order to control annotators' bias, we are planning to use the crowdsourcing platform for semantic similarity annotations for the entire MedSTS dataset, as it has become an easy and inexpensive way to create annotated resources from multiple annotators in a short period of time. Furthermore, we will organize a shared task to invite researchers in the community to tackle with the clinical STS challenge (Rastegar-Mojarad, Liu et al. 2018). We plan to release the MedSTS_ann after manually removing all PHI, and use half as a training dataset and the other half as a testing dataset. Participating teams will be required to sign a Mayo Data Use Agreement to get access to the dataset. They can use the training dataset to build their clinical STS systems. We will release testing dataset later and every team will be allowed to submit 3 runs of their systems. Performance of each system will be evaluated by comparing their system scores against the human scores using the Pearson correlation coefficient as outlined previously in the development of this STS resource, and following SemEval STS shared task precedent.

In addition, we would like to extend our previous system (Afzal, Wang et al. 2016), which was the 3rd participant system in the SemEval 2016 English STS task, to the clinical STS tasks. The system was designed for general English domain. Therefore, we hypothesize that the system could be further improved by incorporating the clinical domain specific features. Recently, deep learning has been prevalently utilized to learn high-level semantic representations (Yan, Yin et al. 2015, Wang, Liu et al. 2018). Further more, we plan to learn word embeddings (Wang, Rastegar-Mojarad et al. 2017) from a large clinical corpus and use those embeddings as features in our previous system for the clinical STS tasks.

# Acknowledgements

This work was made possible by NIGMS R01GM102282, NLM R01LM11934 and NIBIB R01EB19403.